\documentclass[twocolumn,tighten]{aastex62}
\usepackage[utf8x]{inputenc}
\usepackage{amsmath}
\usepackage{amsfonts}
\usepackage{amssymb}
\usepackage{mathrsfs}
\usepackage{bm}
\usepackage{esint} %for varointclockwise
\usepackage{ulem} %for sout

\defcitealias{LDL2019a}{Paper I}

\shorttitle{Long-Lived Eccentricities in Accretion Disks}
\shortauthors{Lee et al.}

\begin{document}
\title{Long-Lived Eccentricities in Accretion Disks}

\correspondingauthor{Wing-Kit Lee}
\email{wklee@northwestern.edu}

\author[0000-0002-5319-3673]{Wing-Kit Lee}
\author[0000-0001-8291-2625]{Adam M. Dempsey}
\author{Yoram Lithwick}
\affiliation{Center for Interdisciplinary Exploration and Research in Astrophysics (CIERA) and Department of Physics and Astronomy, Northwestern University, 2145 Sheridan Road, Evanston, IL 60208, USA}

\begin{abstract}
Accretion disks  can be eccentric: they support $m=1$ modes that are global and slowly precessing.
But whether the modes remain trapped in the disk---and hence are long-lived---depends on conditions at the 
outer edge of the disk.  Here we show that in disks with realistic boundaries, in which the
surface density drops rapidly beyond a given radius, eccentric modes are trapped
and hence long-lived. We focus on pressure-only disks around a central mass,
and show how this result can be understood with the help of a simple second-order WKB
theory. 
We show that the longest lived mode is the zero-node mode in which all of the disk's elliptical
streamlines are aligned, and that this mode decays coherently on the viscous timescale of the disk. 
Hence such a mode, once excited, will live for the lifetime of the disk. It may be
responsible for  asymmetries seen in recent images of protoplanetary disks.
\end{abstract}


\section{Introduction}

Accretion disks are usually assumed to be circular. But for disks in nearly Keplerian potentials, orbits are in general eccentric, and hence the disk as a whole might be eccentric. Eccentric disks are interesting for a variety of reasons; e.g., their distorted shape could be observed, and planets born within such disks would be eccentric, perhaps explaining observed planetary eccentricities.

Eccentric orbits of fluid around a star precess differentially due to the effects of pressure, self-gravity, and non-Keplerian potential components. However,  particular eccentricity profiles may be found such that the  disk as a whole precesses rigidly while maintaining its eccentricity.  
To find such profiles, one may linearize the equation of motion for the (complex)
 eccentricity, in which case the solution is a sequence of normal modes.
 The   ``fundamental'' zero-node mode, i.e., the one in which the eccentricity is never zero, is typically the one of primary interest (e.g., it  lives longest in the presence
 of viscosity).
 
The aforementioned normal mode calculation has been studied by many authors \citep*{1991ApJ...381..259L,2001AJ....121.1776T,Pap2002,2006MNRAS.368.1123G,2008MNRAS.388.1372O,2009MNRAS.400.2090S,2010MNRAS.406.2777L,2016MNRAS.458.3221T}.
In \citet*{LDL2019a} (hereafter \citetalias{LDL2019a}), we solved 
the normal mode problem in disks subject to both pressure and self-gravity forces, and also
explained the numerical results in terms of a simple WKB theory. 
However, an important problem with virtually all previous studies 
is the boundary condition. It is usually assumed that the disk has sharp edges, beyond which
 the surface density drops instantaneously to zero.
But real disks likely have a more gradual drop \citep{1974MNRAS.168..603L}. 
One might worry that normal modes would not be trapped in such a disk, leading to no solutions in which the disk can remain eccentric.

In this paper, we shall show that this worry is unfounded.  Virtually any disk with a realistic
surface density profile will lead to trapped eccentric modes.  
Furthermore, we calculate with numerics and theory the eccentricity profile and precession rate  that occur in disks of arbitrary surface density profiles. Throughout this paper, we focus
on the pressure-only case, i.e., we ignore self-gravity. We do this for simplicity, but also because
even if self-gravity is important in the bulk of the disk, in the outer part where the surface density is small pressure will play the larger role (\citetalias{LDL2019a}). 
We organize this paper as follows. In Section \ref{sec:formulation}, we present the equations of motion
and their WKB formulation.  In Section \ref{sec:diskmodel} we present numerical eigen-solutions, and show how these results may be understood with WKB theory.
In  Section \ref{sec:discussion} we address additional effects, before concluding in 
Section \ref{sec:conclusion}.

\section{Equation of Motion}
\label{sec:formulation}

We consider a two-dimensional gas disk orbiting a central object of mass $M_\star$. The disk is cold and  thin, such that its aspect ratio is much less than one. 
Perturbed variables are assumed to depend on time and azimuth in proportion to $e^{-i\omega t+m\varphi}$. We further set $m=1$, as is the case for eccentric perturbations, in which case
$\omega$ may be identified as the precession frequency.
Using the equations of continuity, momentum, and entropy, one can derive the following eccentricity equation by expressing the fluid variables in terms of eccentricity (\citealt{2006MNRAS.368.1123G}; \citealt{2008MNRAS.388.1372O}; \citealt{2016MNRAS.458.3221T}; see Appendix A of \citetalias{LDL2019a}):
\begin{align}
\label{eq:ecc1_1}
\omega E = \frac{1}{2\Omega r^3 \Sigma}\left\{\frac{d}{dr}\left(\gamma r^3 P \frac{dE}{dr}\right) + r^2 \frac{dP}{dr} E\right\},
\end{align}
where 
$E$ is the complex eccentricity, $\Sigma$, $P$, and $\Omega=\sqrt{GM_*/r^3}$ are the surface density, two-dimensional pressure, and rotational frequency of the gas, respectively, and $\gamma$ is the adiabatic index. The complex eccentricity $E=|E|e^{-i\varpi}$ is used because both the amplitude $|E|$ and the periapse angle $\varpi$ are radial functions. We assume the perturbation is adiabatic\footnote{Corresponding equations for locally-isothermal perturbations (i.e., short cooling time) and vertically-integrated 3D disks can be found in \citet{2016MNRAS.458.3221T}.}, while the equilibrium disk may have non-constant background entropy. 
In deriving Equation \eqref{eq:ecc1_1} we assume the mode frequency $\omega$ is much smaller than the orbital frequency $\Omega$ of the disk, i.e., the mode precesses very slowly.

\subsection{Trapped Modes}
\label{sec:trapping}

\begin{figure}[t]
	\centering
	\includegraphics[trim={0.cm 1.1cm 1cm 2cm}, clip, 
	width=\linewidth]{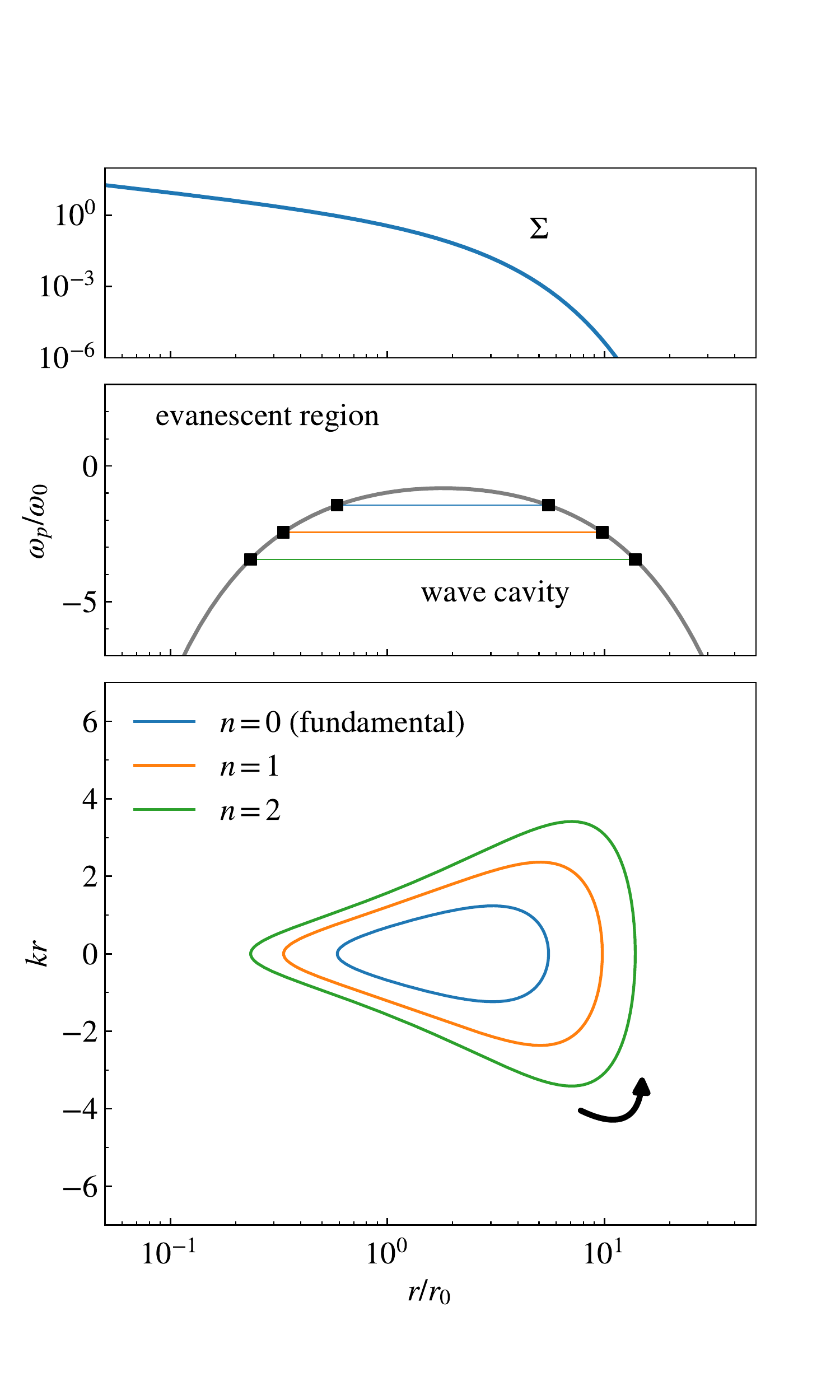}
	\caption{A disk with background profiles $T \propto r^{-1/2}$ and $\Sigma = r^{-1} e^{-r/r_0}$ is considered. (Top) The surface density profile in arbitrary unit. (Middle) A frequency diagram. The blue curve shows $\omega_p$ normalized by $\omega_0$  (which is defined in Equation \eqref{eq:omega0}). The numerical spectrum of the highest 3 frequencies is shown as the horizontal lines, in which their radial extents represent the wave cavity. Turning points are denoted by black squares.  (Bottom) The dispersion relation map (DRM), showing contours of constant $\omega$ for the three modes in the middle panel.  The  arrow marks the direction of propagation. The phase-space area enclosed by each mode is an odd integral multiple of $\pi$ (i.e., quantum condition in Equation \eqref{eq:est1_0}).}
	\label{fig:frequencyDRM}
\end{figure}

Equation \eqref{eq:ecc1_1} may be cast into a more  transparent form by  
transforming  variables from $E$ to 
\begin{equation}
y=(r^3 P)^{1/2} E \ ,
\label{eq:ydef}
\end{equation}
which leads to an equation with no single-derivative ($dy/dr$) term:
\begin{align}
\label{eq:ecc1_3}
\frac{d^2 y}{dr^2} + \frac{2\Omega}{c^2}\left(\omega_p(r)-\omega\right) y &= 0,
\end{align}
where  
\begin{equation}
c=\sqrt{\gamma P/\Sigma} \nonumber
\end{equation}
 is the sound speed, and the ``effective potential" $\omega_p$ is given by
\begin{align}
\label{eq:ecc1_5}
\omega_p(r) = -\frac{c^2}{2\Omega}\left[(r^3 P)^{-1/2}\frac{d^2}{dr^2}(r^3 P)^{1/2}-\frac{1}{\gamma rP}\frac{dP}{dr}\right].
\end{align}

Equation (\ref{eq:ecc1_3}) is particularly simple to analyze because of its similarity
to  Schr\"{o}dinger's equation, with
 $\omega_p$ playing the role of the potential and $\omega$ the role  of the energy\footnote{
 Unlike Schr\"{o}dinger's equation, Equation (\ref{eq:ecc1_3}) has a spatially-variable coefficient
 ($2\Omega/c^2$)
 multiplying its energy term. That may be fixed by transforming co-ordinates 
  \citep[e.g.,][]{2008MNRAS.388.1372O,2009MNRAS.400.2090S}, but we do not do so.}.
It is apparent that $y$ is wave-like where $\omega_p>\omega$ and evanescent where $\omega_p<\omega$, with
  turning points at $\omega_p=\omega$. 

\begin{figure}[t]
	\includegraphics[trim={0.5cm 0.7cm 0.5cm 0},clip,width=0.48\textwidth]{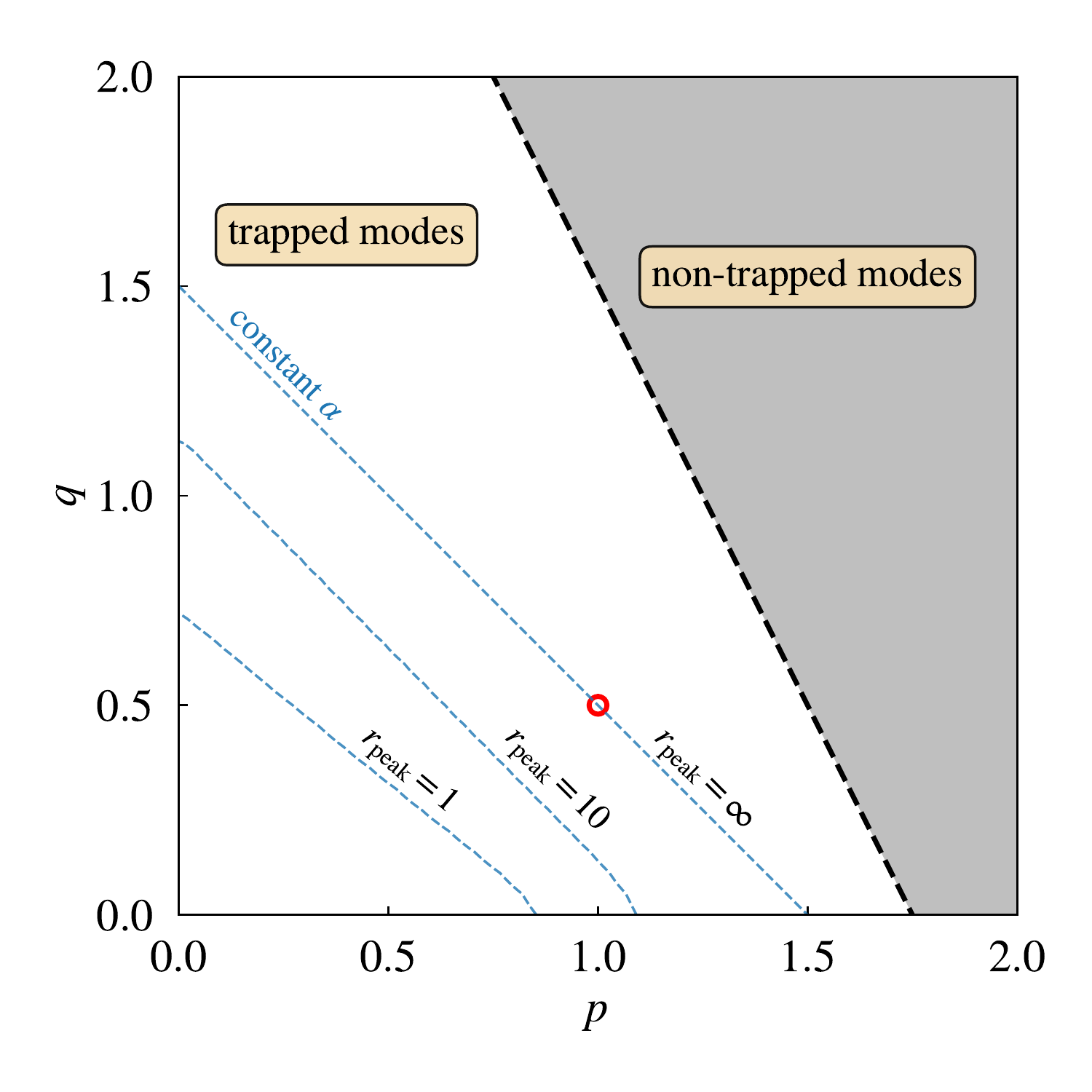}
	\caption{Regions of interest within parameter-space, where the axes $p$ and $q$ denote
	the power-laws of the background surface density and temperature profiles. 
	The white region shows where modes are trapped, i.e., where $\omega_p$ takes on an inverted-U shape. The dashed black line bounding the white region is the line $2p+q=7/2$, beyond which $\omega_p$ rises outwards.
	For constant $\alpha$ self-similarly evolving disks, $p$ and $q$ are constrained to the line $q=3/2-p$; the red circle shows the fiducial case $(p,q)=(1,1/2)$ considered  in this work. 
       The blue dashed  lines that are labelled by $r_{\rm peak}$ (one of which is identical to
       the constant $\alpha$ line) show where the eccentricity of the fundamental mode
       peaks, as discussed in Section \ref{sec:asymptoticsolution}.}
	\label{fig:parameterspace}
\end{figure}

For realistic disk profiles, the ``potential'' $\omega_p$ almost always has an inverted-U shape.  
 As a result, eccentric modes are trapped by the peak of the potential, and their character will not depend on what happens far from the peak---a central result of this paper.
For example, we plot $\omega_p$ in the middle panel of Figure \ref{fig:frequencyDRM} for our ``fiducial case'': a disk with the temperature and surface density profiles $T \propto r^{-1/2}$ and $\Sigma = r^{-1} e^{-r}$ (Figure \ref{fig:frequencyDRM}, top panel), respectively.
More generally, motivated by the self-similar  solutions of \citet{1974MNRAS.168..603L}, we  consider  profiles of the form
\begin{eqnarray}
\label{eq:profile1a}
T(r)&\propto& r^{-q} \label{eq:temp} \ , \\
\label{eq:profile1b}
\Sigma(r)&\propto& r^{-p}e^{-(r/r_0)^\xi} \ , \label{eq:sig}
\end{eqnarray}
i.e., 
with a cutoff on the surface density at $r\gtrsim r_0$, and two free parameters, $p$ and $q$. For simplicity, we set $\xi$ in the above equations to $\xi=2-p$, as
is the case for a self-similarly evolving disk that has  power-law viscosity\footnote{If the disk also has an $\alpha$-viscosity with constant $\alpha$, then $q=3/2 -p$. \label{foot:alph}}.
The resulting $\omega_p$ is a sum of three power-laws (given explicitly in Equation \eqref{eq:ompp}). 
In Figure \ref{fig:parameterspace} the white region shows where, in the $p$-$q$ 
plane,  $\omega_p$ has an inverted-U shape, and so can trap modes.
In the grey region $\omega_p$ rises continually outwards, and so modes cannot
be trapped. For the remainder of this paper, we consider only the white region because
it encompasses most typically assumed values for $p$ and $q$.

\subsection{Second-order WKB Theory}

An  advantage of analyzing Equation \eqref{eq:ecc1_3} rather than \eqref{eq:ecc1_1}  is that
one may derive a second-order-accurate WKB dispersion relation by replacing $d/dr \rightarrow ik$ \citep{2007AN....328..273G}:
\begin{align}
\label{eq:ecc1_4b}
\omega = \omega_p - \frac{c^2}{2\Omega} k^2,
\end{align}
where $k$ is the radial wavenumber. We consider Equation \eqref{eq:ecc1_4b} as a second-order dispersion relation because $\omega_p$ is two orders of $kr$ smaller than the leading term \citepalias{LDL2019a}.\footnote{Equation (\ref{eq:ecc1_1})  admits a conserved, i.e., spatially constant, quantity 
$F=-i{m\pi \gamma \over 2}(y^*{dy\over dr}-y{dy^*\over dr})$, that is equal  to the  angular momentum flux. In WKB, $F= m \pi \gamma r^3Pk|E|^2$ \citep{1979ApJ...233..857G}.}

Following \citetalias{LDL2019a}, the dispersion relation can be analyzed using a frequency diagram
(Figure \ref{fig:frequencyDRM}, middle panel) and  dispersion relation map (DRM; Figure \ref{fig:frequencyDRM}, bottom panel), which plots contours of constant $\omega$ in the $r$-$kr$ plane. 
The frequency diagram shows the trapped modes. The mode with highest frequency is the zero-node
fundamental mode, and those with more negative frequency have increasing number of nodes. 
Note that in this example (and as is typically true in the absence of self-gravity forces) frequencies are negative, meaning modes are retrograde. 
From the DRM, we see how a trapped wave refracts as it propagates, transitioning from outwardly-propagating (in the lower half of the figure) to inwardly propagating (top half) at its turning point ($k=0$). Its group velocity is $v_g = (\partial \omega/\partial k)_r = - kc^2/\Omega$.

In order for a trapped wave to represent a standing mode, its complex phase must change by 
an integral multiple of $2\pi$ over the course of a loop in the DRM.  Accounting for the phase 
change at turning points, that implies that standing modes must satisfy
\begin{align}
\label{eq:est1_0}
\varointclockwise k dr = (2n+1)\pi = \pi, 3\pi, 5\pi, \cdots,
\end{align}
where the path-integral is taken in the clockwise direction in phase-space (according to the group velocity), and $n=0,1,2,\cdots$ labels the number of nodes (\citealt*{1977ApJ...212..645M}; \citealt{1990ApJ...358..495S}; \citetalias{LDL2019a}). Equation \eqref{eq:est1_0} allows us to determine the mode frequency from theory, after inserting
$k(r;\omega)$ from the dispersion relation. 

\section{Solutions}
\label{sec:diskmodel}

\subsection{Numerical Solutions}

\begin{figure*}[t]
	\centering
	\includegraphics[trim={0.cm 0.cm 0.cm 0},clip,width=\linewidth]{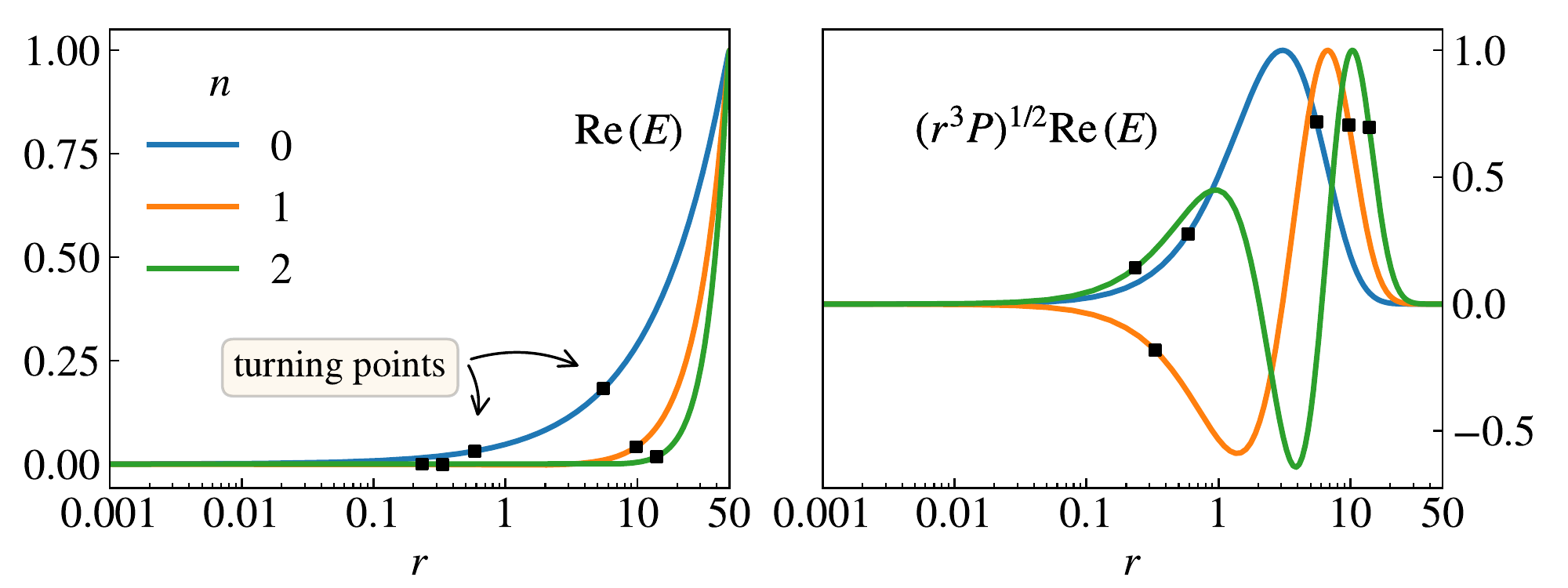}
	\caption{Plot of first 3 eccentric modes with fewest nodes ($n$). The left and right panels show the real parts of $E$ and $y=(r^3 P)^{1/2} E$, respectively (both are normalized to their peak values, and the imaginary part vanishes.). The outer computational boundary is at $50$. The black squares mark the turning points taken from Figure \ref{fig:frequencyDRM}.}
	\label{fig:pressureeigfuncjan19}
\end{figure*}

We solve Equation (\ref{eq:ecc1_1}) numerically; in subsequent subsections 
we shall explain the results with second-order WKB theory.
For the background state, we adopt Equations (\ref{eq:temp})--(\ref{eq:sig}), 
with $p=1$, $\xi=1$, and $q=1/2$, and explore other values for these parameters below.
We set the 2D adiabatic index to $\gamma=3/2$. 
Using $P=\Sigma c^2/\gamma$
in Equation (\ref{eq:ecc1_1})   implies that its  right-hand side is proportional
to 
\begin{eqnarray}
\label{eq:omega0}
\omega_0\equiv {c_0^2\over 2\Omega_0 r_0^2} \ ,
\end{eqnarray}
where $c_0$ and $\Omega_0$ are evaluated at the density cutoff $r_0$.
The quantity $\omega_0$  is the characteristic frequency that determines the precession
 frequency of the modes, aside from dimensionless constants. 
 We can write $\omega_0 \sim h_0^2 \Omega_0$ where $h_0$ is the aspect ratio of the disk at the outer cut-off, i.e., it is slower than the orbital frequency at the disk's outer cutoff by the square of the aspect ratio there. Henceforth we shall measure mode frequencies in units of $\omega_0$ (i.e., set $\omega_0\rightarrow 1$). We choose our length unit to be $r_0$, i.e., set $r_0\rightarrow 1$.

For boundary conditions, the inner and outer disk edges are assumed to be free surfaces (i.e., zero Lagrangian pressure), such that the following boundary condition holds (\citealt{2008MNRAS.388.1372O}, \citealt{2016MNRAS.458.3221T}, \citetalias{LDL2019a}):
\begin{align}
\label{eq:ecc1_2}
dE/dr = 0.
\end{align}
The inner and outer radii of the disk are $r_{\rm min}=10^{-4}$ and $r_{\rm max}=50$, respectively.

We solve the boundary-eigenvalue problem of  Equations \eqref{eq:ecc1_1} and \eqref{eq:ecc1_2}
with a finite difference method (see \citetalias{LDL2019a} for details) and a Chebyshev spectral method \citep{SpectralMethods}. In both cases, we construct a square matrix and solve for the eigenvalues and eigenfunctions using the LAPACK library. The two solution methods give essentially the same results, and we have checked that the results have converged with respect to the number of grid points. 

\subsection{Eigenfunctions}

Figure \ref{fig:pressureeigfuncjan19} shows the numerical eigenfunctions for the first few modes. 
From the right-hand panel, which shows the scaled eccentricity $y$ (Equation \eqref{eq:ydef}), 
we see that the oscillatory and evanescent regions  predicted from the frequency diagram (Figure \ref{fig:frequencyDRM}) match up with those of the solutions.  The left-hand panel shows
the unscaled eccentricity $E$, which rises outwards, as explained in further
detail in Section \ref{sec:asymptoticsolution}.

\subsection{Eigenvalues}
\label{sec:estimate_model}

\begin{figure}[t]
	\centering
	\includegraphics[trim={0.5cm 0.7cm 0.5cm 0},clip,width=\linewidth]{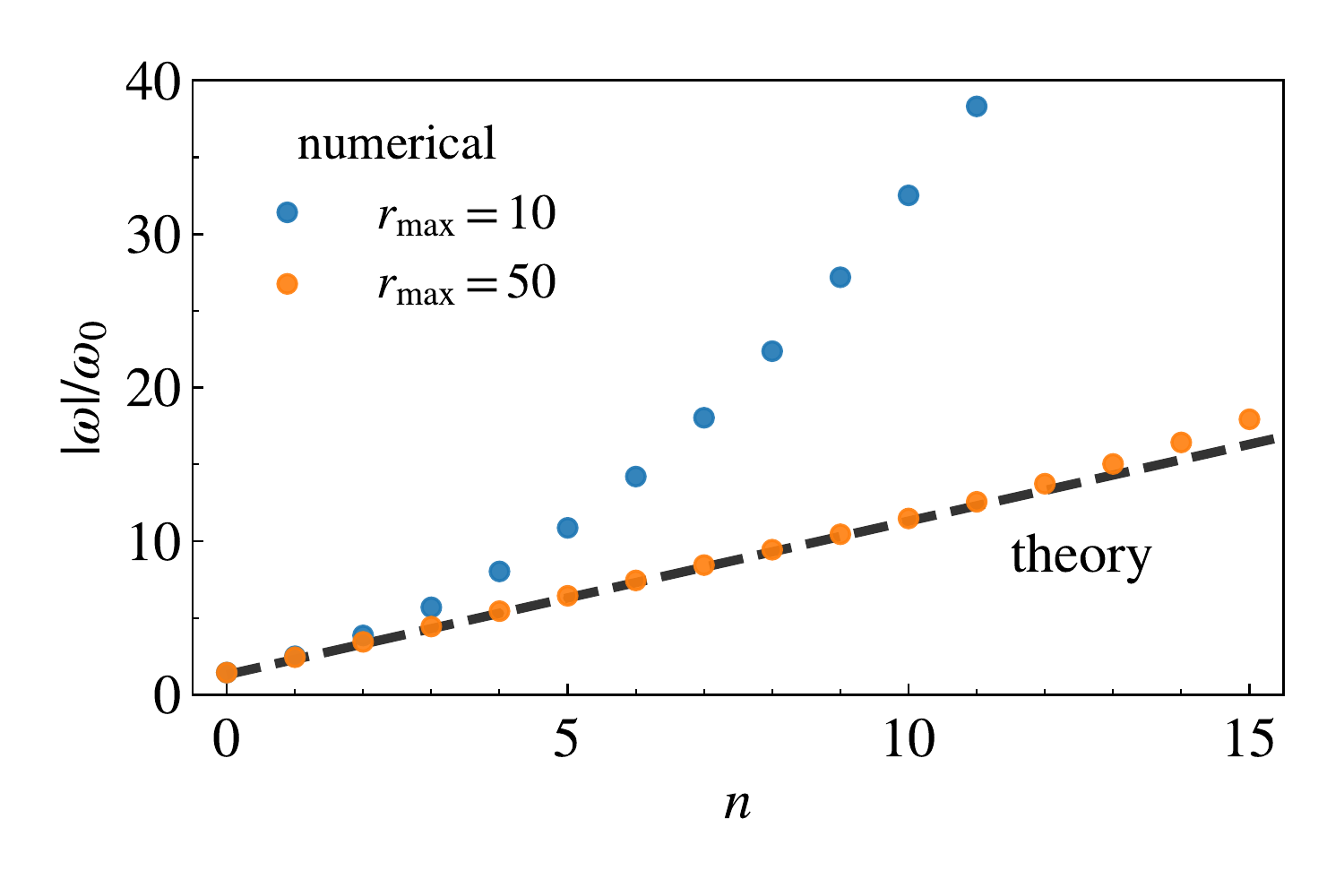}
	\caption{The mode frequencies $\omega$ against number of nodes $n$ are shown for a disk with $(p,\xi,q) = (1,1,1/2)$. The orange
	dots   mark the numerically computed $\omega$ for $r_{\rm max}=50$; the
	blue dots are for a case with lower $r_{\rm max}$, in which case the boundary
	gives rise to incorrect results at high $n$. The black dashed curve is the theoretical estimate using the WKB theory in Equation \eqref{eq:est1_8}.
	}
	\label{fig:pressureomeganjan19}
\end{figure}

In Figure \ref{fig:pressureomeganjan19}, the orange dots show the numerically calculated
eigenfrequencies up to $n=15$. Also shown, as blue dots, is the result from reducing
the outer boundary from $r_{\rm max}=50$ to 10. For small $n$ the two sets agree, 
while for larger $n$ they diverge. The reason for the divergence is that,
in the lower $r_{\rm max}$ simulation, 
the turning points no longer fit in the simulation domain when $n\gtrsim 3$
 (Figure \ref{fig:pressureeigfuncjan19}, right panel).  
 
We may calculate the eigenfrequencies theoretically by 
inserting $k(r;\omega)$ from the dispersion relation (Equation \ref{eq:ecc1_4b}) into the quantum condition (Equation \ref{eq:est1_0}). For the fiducial case  $(p,\xi,q)=(1,1,1/2)$ (red circle in Figure \ref{fig:parameterspace}), the integral can be performed analytically, yielding a harmonic-oscillator-like form:
 \begin{align}
\label{eq:est1_8}
\omega = -\left(n+\frac{1}{2}\right) + \omega_{p,\rm peak},
\end{align}
where $\omega_{p,\rm peak}=-(B+2\sqrt{AC}) \simeq -0.81$ is the peak value of $\omega_p$. Figure \ref{fig:pressureomeganjan19} shows that the WKB formula above provides an excellent match to the numerical eigenvalues. In fact, it might appear that the agreement is too good at small $n$, given that we are making the WKB approximation, which one might expect to fail for the low $n$ modes. The reason is that,  near the peak of $\omega_p$,  the exact equation (Equation \ref{eq:ecc1_3}) is similar to the equation for a harmonic oscillator, for which the WKB solution is exact.

We now consider the more general case where 
$p$ and $q$ are arbitrary (but $\xi=2-p$), proceeding approximately. 
The quantum condition reads 
\begin{eqnarray}
(2n+1)\pi&=& \int_{r_-}^{r_+}   r^{(2q-3)/4}\left(\omega_p-\omega\right)^{1/2} dr  \\
&\sim& r_+^{(2q+1)/4}(-\omega)^{1/2}
\end{eqnarray}
where $r_\pm$ are the turning points, and in the latter expression we dropped
$\omega_p$ because it becomes small (in magnitude) relative to $\omega$ far
from the turning points.
To estimate $r_+$ we set $\omega=\omega_p$, where for $\omega_p$ we
use its dominant piece, i.e., the last term in Equation (\ref{eq:ompp}). We find
\begin{equation}
|\omega| \sim n^{1/\delta} \ ,
\end{equation}
where $\delta = (2q+1)/(14-8p-4q)+1/2$ and we omit order-unity coefficients.

\subsection{Behavior of Eigenfunction at large radii}
\label{sec:asymptoticsolution}

\begin{figure}[t]
	\centering
	\includegraphics[trim={0.6cm 0.4cm 0.8cm 0},clip,width=\linewidth]{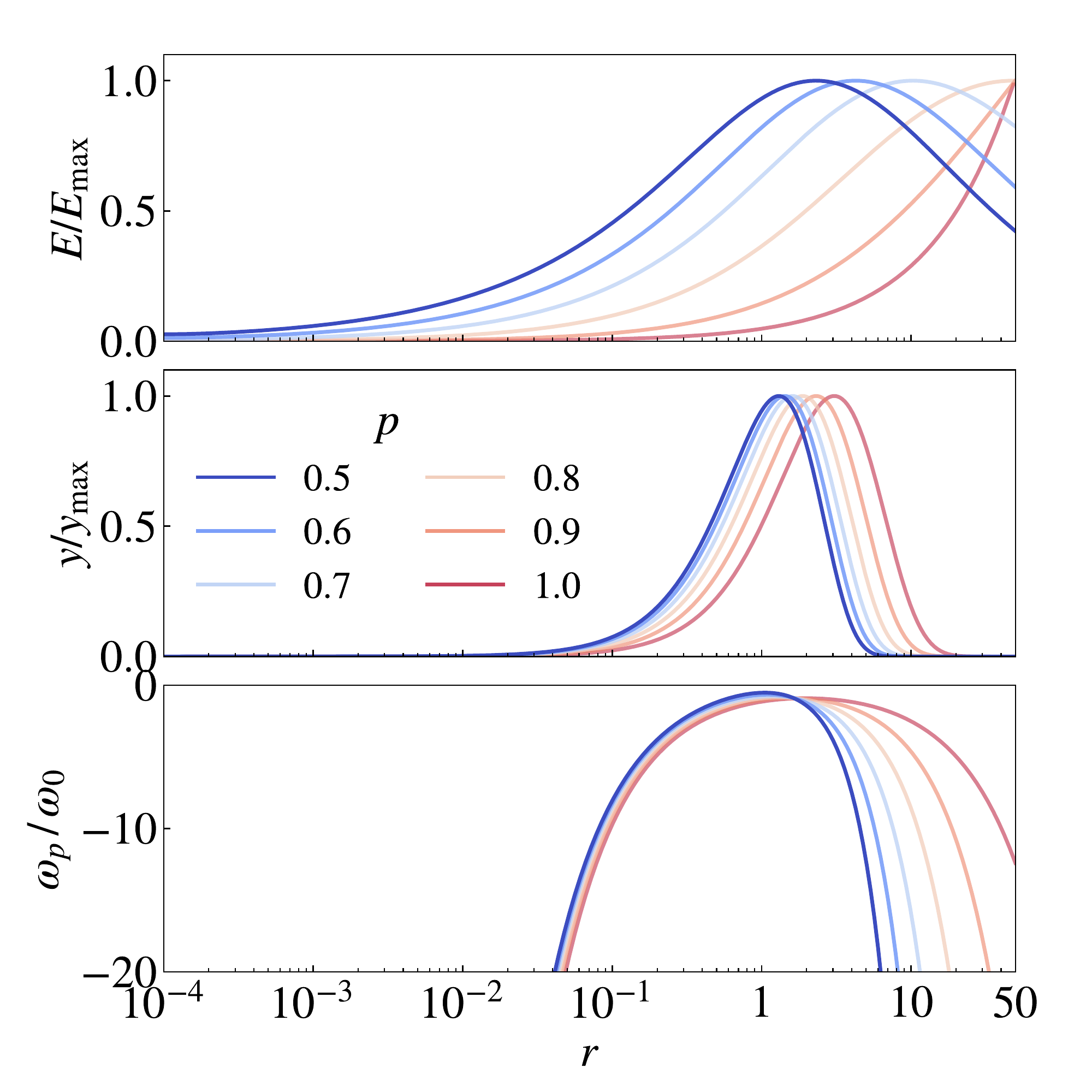}
	\caption{Comparison of the fundamental modes for disks with different $p$ (and $\xi=2-p$) and fixed $q=1/2$. In the middle panel, we show that all modes are trapped (i.e., $y$ decreases to zero near the edges). The width of the eigenfunctions roughly coincide with that of the precession rates shown in the bottom panel.}
	\label{fig:compare_p_fixed_q}
\end{figure}

Figure \ref{fig:pressureeigfuncjan19} shows that at $r\gtrsim 1$ 
the eccentricity continues to rise outwards. Since the behavior
near $r\sim 1$ is potentially observable, we 
examine it here in more detail. 
  Figure \ref{fig:compare_p_fixed_q} shows the fundamental modes 
  for a set of background profiles with different $p$'s. 
 Sometimes the eccentricity is peaked within the disk, while sometimes 
   it rises continuously outwards.
   For more general values of $p$ and $q$, we have determined numerically 
   where the eccentricity of the fundamental mode peaks.  Our results are
   shown as blue dashed lines in Figure  \ref{fig:parameterspace}, labeled
   by $r_{\rm peak}$ (the radius where they peak).
   Curiously, beyond the constant $\alpha$ line the eccentricity peaks at infinite
   $r$; otherwise, it peaks further in.

We may  understand this behavior from WKB theory.
 The solution beyond the outer turning point  is a decaying exponential function for $y$ \citep[e.g.,][]{BenderOrszag1999}, which leads to
\begin{align}
\label{eq:asym1_1}
E \sim (r^3 P)^{-1/2}\exp\left[-S(r)\right],
\end{align}
where $S=\int^r_{r_+} \sqrt{-k^2(s)}ds$ and we assume $r_+ \ll  r$. Inserting the expression for $k$ from the dispersion relation, and Taylor
expanding in small $1/r$ yields $\sqrt{-k^2}\approx C^{1/2}r^{\xi-1}\left[ 1+{1\over 2C}(Br^{-\xi}+\omega r^{q-2\xi+1/2}) \right]$.
Note that the last term is subdominant because
 $q-2\xi+1/2<0$ in the white ``trapped mode'' zone in Figure \ref{fig:parameterspace}.
 Therefore, we get
\begin{equation}
	E\sim r^{{p+q-3\over 2}-\frac{B}{2-p}} \exp\left[{-{\omega r^{p+q-3/2}\over (2-p)(p+q-3/2)}}\right].
\end{equation}
We see that for $p+q-3/2>0$ (i.e., beyond the constant $\alpha$ line), the eccentricity rises inexorably outwards, while in the opposite limit it is peaked in the disk, confirming
the numerical result.

\section{Discussion}
\label{sec:discussion}

\subsection{Viscous Damping}
\label{sec:viscosity}
To study  eccentricity damping, 
we consider the effects of bulk viscosity, following
 \citet{2006MNRAS.368.1123G}.
 We ignore shear viscosity because it may cause overstability \citep{2001MNRAS.325..231O}. 
The linearized equation of motion with bulk viscosity is obtained by replacing the adiabatic index
$\gamma$ by $\gamma+i\alpha$ in Equation (\ref{eq:ecc1_1}), 
where $\alpha$ is the Shakura-Sunyaev parameter (but for bulk viscosity rather than 
the more common shear viscosity). 
We solve  this modified equation numerically in  the same way as before, 
 and obtain the damping rate as the imaginary part of the eigenfrequency $\Gamma = -{\rm Im}\,\omega$.

We find that the numerical damping rate is of order of $\alpha |\omega|$. This can be understood by considering an integral relation from Equation \eqref{eq:ecc1_1}
\citep[][]{2006MNRAS.368.1123G,2010MNRAS.406.2777L,2016MNRAS.458.3221T}:
\begin{align}
\label{eq:damping1_1}
\Gamma =  \frac{\int \alpha r^3 P \big|dE/dr\big|^2 dr}{\int 2\Omega r^3 \Sigma |E|^2 dr} \sim \frac{\int \alpha \big|dy/dr\big|^2 dr}{\int (2\Omega/c^2) |y|^2 dr},
\end{align}
where the absolute sign denotes the amplitudes. The last approximation is justified because the numerator is dominated by the oscillatory region within the wave cavity (Figure \ref{fig:pressureeigfuncjan19}). Under the WKB approximation, the integrals can be estimated as $\Gamma \sim \alpha k^2 c^2 / 2\Omega \sim \alpha |\omega|$, consistent with the numerical result. 

This damping rate $(\Gamma\sim \alpha|\omega|)$ is slow.  In particular, since $\omega\sim \omega_0$, the damping time $\Gamma^{-1}$ is comparable to the viscous
time of the disk as a whole.  Therefore, once such a mode is excited, it will live for
the lifetime of the disk. 

\subsection{Three-dimensional Effects}

We have assumed throughout this paper that the disk is two-dimensional. But
\citet{2008MNRAS.388.1372O} pointed out that in 3D disks   an extra precession term needs to be included because  a fluid parcel cannot maintain vertical hydrostatic equilibrium while in an eccentric orbit. He derived an equation analogous to Equation (\ref{eq:ecc1_1}) valid for 3D disks. 
We have analyzed that equation in the same way as we analyzed Equation \eqref{eq:ecc1_1}, 
but we do not present details here---primarily because in order to derive a reduced
1D equation (such as Equation \eqref{eq:ecc1_1}), one must make the questionable assumption that the eccentricity is independent of height. Nonetheless, the result is very similar to the one found in this paper, i.e., there is an $\omega_p$ function that traps the modes.  But there is one important difference, which is that  in the inner disk the sign of the dominant term of $\omega_p$ switches to positive, which can potentially remove the inner turning point (the outer turning point is unaffected), and hence the modes must rely on the reflection at the disk inner edge in order to remain trapped \citep[e.g.,][]{2018ApJ...857..135M}. 

\section{Conclusion}
\label{sec:conclusion}

\begin{enumerate}
    \item We demonstrate that typical accretion disks with a realistic outer density drop  support trapped eccentric modes. We use the second-order WKB theory developed in \citetalias{LDL2019a} to explain the different features, such as the wave cavity and the eccentricity in the outer part of the disk.
    \item Each normal mode solution corresponds to a rigidly precessing eccentric pattern. This is a balance between the pressure precession effects caused by the axisymmetric and non-axisymmetric components of the disk. We find that, instead of the ``test-particle precession rate" (i.e., the difference between orbital and epicyclic frequencies $\dot{\varpi} = \Omega-\kappa$) in celestial mechanics, the $\omega_p$ function gives the correct behavior of eccentricity of a gas disk. Previous results based on the leading-order WKB dispersion relation \citep[e.g.,][]{Pap2002,2003ApJ...585.1024G} may require extra examination.
    
	\item We find that trapped eccentric modes are standing waves with a discrete spectrum (Figure \ref{fig:frequencyDRM}). The fundamental mode has zero radial nodes and the least negative frequency. It has the slowest damping rate when viscous damping is considered (Section \ref{sec:viscosity}).
	
	\item The eccentricity of a mode in the outer disk is evanescent and does not carry angular momentum. It can also be explained by the WKB theory (Section \ref{sec:asymptoticsolution}). The eccentricity of a disk with $p+q < 3/2$ peaks inside the disk rather than at infinity (Figure \ref{fig:parameterspace}). 
	\item The trapped modes are not affected by the boundary conditions as long as the artificial computational boundary (i.e., $r_{\rm max}$) is far from the turning point (Figure \ref{fig:pressureomeganjan19}). 
\end{enumerate}

Although we have shown that eccentric modes can live for a long time, we have not addressed the question of how  such modes are excited. Some possibilities are gravitational excitation
by a planet or star, or an internal instability. 
 
Finally, we note that a long-lived eccentric disk can possibly be detected directly. 
Some lopsided disks have been detected \citep[e.g.,][]{2018ApJ...860..124D,2018ApJ...869L..41A}, and a variety of mechanisms proposed
to explain them  \citep[e.g.,][]{2012ApJ...760..119H,2013A&A...553L...3A,2013ApJ...775...17L,2015ApJ...798L..25M,2016MNRAS.458.3918Z}. But whether the lopsidedness might be due to the disk being eccentric is an intriguing possibility.

\acknowledgments

WKL thanks Kenny L.S. Yip for checking Equation \eqref{eq:est1_8} using contour integration.  Y.L. acknowledges NASA grant  NNX14AD21G and NSF grant AST-1352369.

\appendix

\section{$\omega_p$ in  model disk}
\label{sec:effpotential}
For a disk model given by Equations (\ref{eq:profile1a})--(\ref{eq:profile1b}) with general parameters $(p,q,\xi)$, 
the $\omega_p$ function in Equation \eqref{eq:ecc1_5} is
given by
\begin{align}
\omega_p = - r^{-q-1/2} (A + B r^{\xi} + C r^{2\xi})  \ .
\label{eq:ompp}
\end{align}
The coefficients are given by 
\begin{align}
\label{eq:app2_5}
A &= \frac{3}{4}+\frac{1}{4}(p+q)^2 + (p+q)\left(\frac{1}{\gamma}-1\right), \\
B &= \frac{\xi}{2}\left(p+q-\xi-2+\frac{2}{\gamma}\right),\\ 
C &= \xi^2/4,
\end{align}
when units are set by $\omega_0=1$ and $r_0=1$. 
For the fiducial case ($p=2-\xi=1$, $q=1/2$, and $\gamma=3/2$),  $A=13/16$, $B = -1/12$, and $C=1/4$.

\bibliography{disk3,pressure2,observations,sdwtheory}
\end{document}